\documentstyle[epsfig,12pt, english, float]{article}

\begin{document}

\begin{abstract}
\noindent Velocity measurements of wind blowing near the North Sea border of Northern Germany and velocity measurements under local isotropic conditions of a turbulent wake behind a cylinder are compared. It is shown that wind gusts -- measured by means of velocity increments -- do show similar statistics to the laboratory data, if they are conditioned on an averaged wind speed value.\\
Clear differences between the laboratory data and the atmospheric wind velocity measurement are found for the waiting time statistics between successive gusts above a certain threshold of interest.
\end{abstract}

\section{Introduction}

The occurances of gusts represent a great problem affecting a wide 
field of 
technical applications as for instance the construction of Wind 
Energy Converters \cite{altemark} as well as the construction of 
high buildings in general \cite{fordham}.
But also regarding the landing- and take-off-operations at airports
the importance of the statistics of gusts becomes evident 
\cite{manasseh}. Nevertheless {\it gusts} are not uniquely defined. 
In \cite{hau} or \cite{book} gusts are taken as an additional structure besides the permanent, fluctuating {\it turbulence} $u(t)$ around the mean wind $\bar{u}(t)$ of the wind 
velocity field. More often wind gusts are characterized by {\it extrem} velocity differences 
during  short time intervals ranging typically between some seconds up to a minute. But 
these characterisations are still vague.\\
In order to examine wind gusts in a statistical way 
we use a data set of a turbulent wind velocity field recorded near the coastline of the Northsea at 
the FH Ostfriesland in Emden (Germany). The velocity was measured by means of an ultrasonic anemometer at $20\;m$ height. The sampling 
frequency was $4\;Hz$. The measuring period took about one year (1997-1998). 
After careful investigation of the quality of the data we examine a representative 275-hour-excerpt of October 1997 and focus on the velocity 
component in direction of the mean wind (See also for a first brief report on the analysis of these data \cite{dewi}). \\
Our main interest is to explore how atmospheric turbulence is related to the well known local isotropic and stationary laboratory one. Therefore we use a simple measure for gusts and compare both velocity fields by means of their probabilistic features. The laboratory data we use here was recorded in a wind tunnel behind a cylinder with a Reynoldsnumber of $Re\approx 30000$ (see \cite{Stefan}).
In a second step we examine the waiting time distributions of successive wind gusts to resolve also their time structure. 

\section{Probabilistic description of wind gusts}

\begin{figure}[H]
  \begin{center}
    \epsfig{file=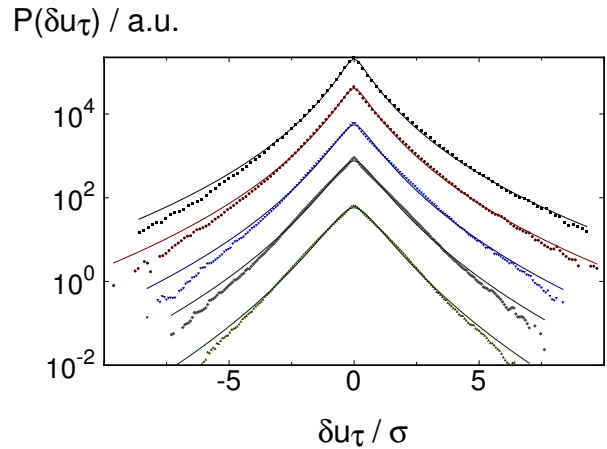, width=8.0cm}
  \end{center}
 	\caption{\it The pdfs of the unconditioned windincrements for $\tau$ being 	$0.008\;T$, $0.03\;T$, $0.2\;T$, $0.95\;T$ and $61\;T$ (full symbols from the top to 	the bottom) are drawn in. The pdfs are shifted against each other for a clearer 	presentation and the respective fitfunctions according to eq. (\ref{c7}) are 	shown as solid curves.}
 	\label{windpdf}
\end{figure}

\begin{figure}[H]
  \begin{center}
    \epsfig{file=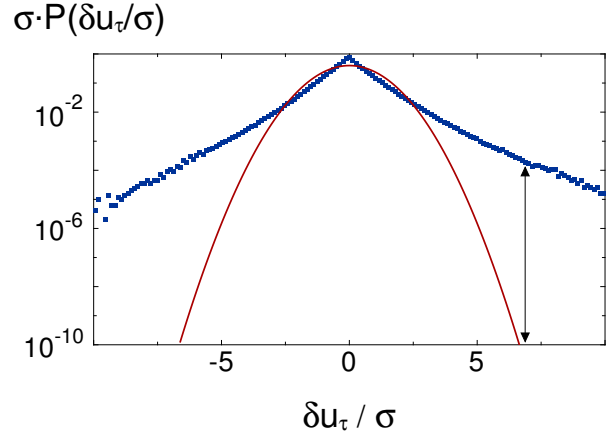, width=8.0cm}
  \end{center}
 	\caption{\it The distribution of the windincrements for $\tau=4s$ is represented by the 	squares, a Gaussian distribution with the same standard deviation $\sigma$ by the 	solid line (parabola due to the semilogarithmic presentation). Both distributions are 	normalized with $\sigma=0.8\;m/s$.}
 	\label{gauss}
\end{figure}

\begin{figure}[H]
  \begin{center}
    \epsfig{file=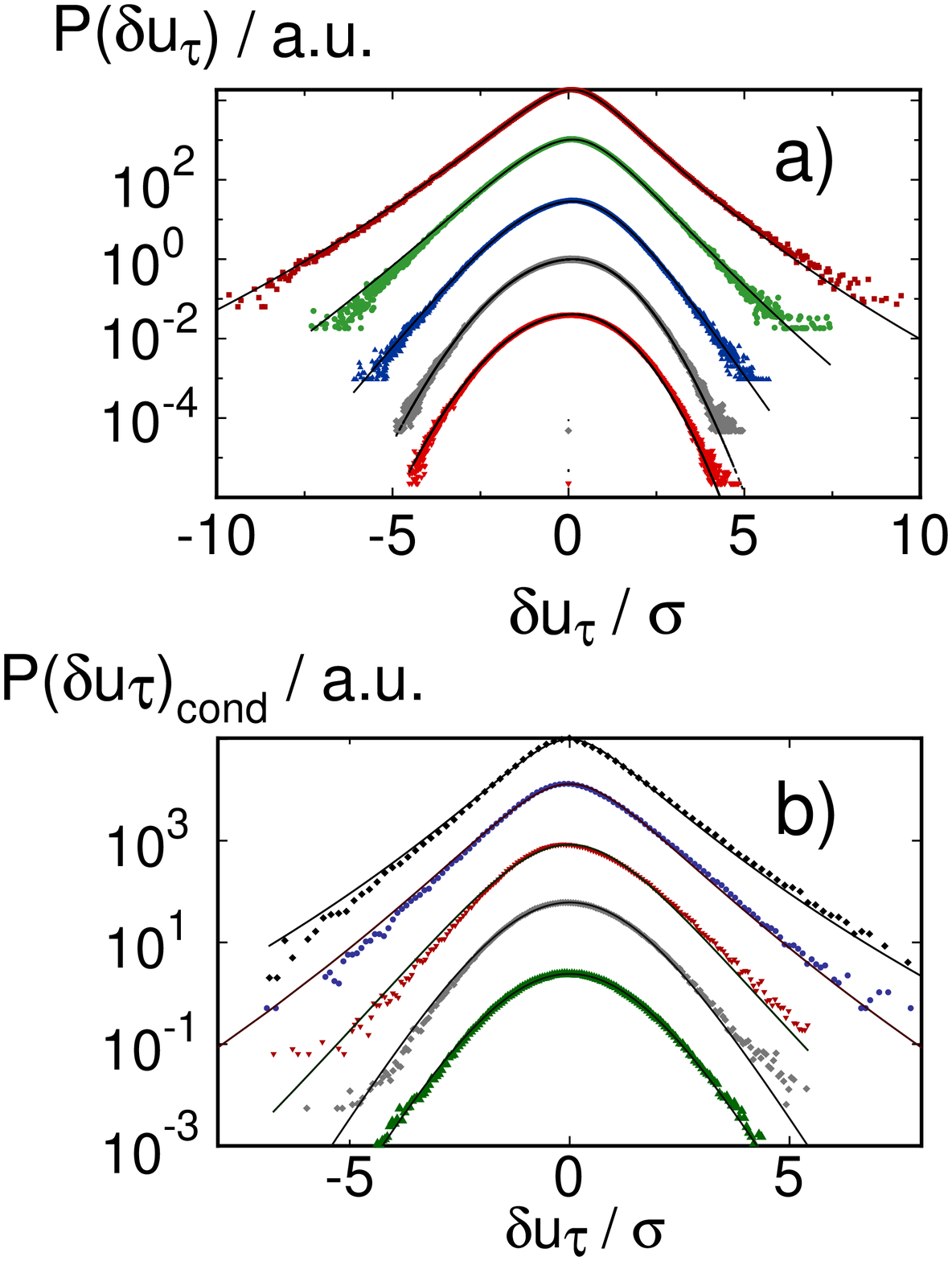, width=8.0cm}
  \end{center}
 	\caption{\it In a) the symbols represent the pdfs $P(\delta u_{\tau})$ of the laboratory 	increments for different values of $\tau$. From the 	top to the bottom $\tau$ takes the values: $0.005\;T$, $0.02\;T$, 	$0.17\;T$, 	$0.67\;T$ and $1.35\;T$. In b) the conditioned pdfs are presented, 	here $\tau$ is $0.008\;T$, $0.03\;T$, $0.2\;T$, $0.95\;T$ and $1.9\;T$. The mean 	wind interval on which the increments are conditioned is $[4.5\;;5.6]\;m/s$. In both 	cases the solid lines are the corresponding fit-functions 	according to eq. (\ref{c7}). The distributions and their fits are shifted against 	each other for a clearer presentation.}
 	\label{Vergleich}
\end{figure}

The wind field is known to exhibit a high degree of turbulence with 
Reynoldsnumbers $Re$ of about $10^{7}$.  The wind velocity $U(t)$ is commonly (see \cite{hau}) expressed as the sum of the mean velocity 
$\bar{u}(t)$ and fluctuations $u(t)$ around it. For the mean velocity $\bar{u}(t)$ a ten minute average is used:
\begin{equation}
	U\;=\;\bar{u}+u \;\;\;	.
	\label{notation}
\end{equation}
Here we concentrate on the velocity components in the mean velocity direction. The greater the fluctuation values $u$ the more turbulent the wind field becomes. In windenergy research this is often expressed by the {\it turbulence degree} $t_{i}$ which is defined as the standard deviation $\sigma$ in relation to the mean velocity $\bar{u}$:
\begin{equation}
	t_{i}\;=\;\frac{\sigma}{\bar{u}}\;\;\; .
	\label{turbgrad}
\end{equation}
\noindent Nevertheless the value of $t_{i}$ does not contain any dynamical or time-resolved information about the fluctuation field itself. \\
\begin{figure}[H]
  \begin{center}
    \epsfig{file=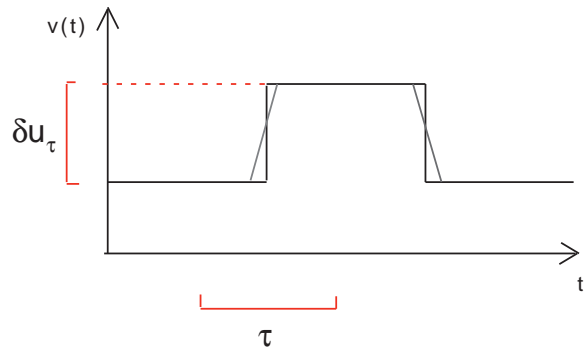, width=8.0cm}
  \end{center}
 	\caption{\it Idealized representation of a wind gust.}
 	\label{trapez}
\end{figure}
To achieve a deeper understanding of wind gusts -- as a result of the fluctuating wind field -- we investigate in how far wind gusts are related to the well known features of small scale turbulence. To this end we will perform an analogous analysis for our wind data and data from a turbulent wake flow behind a cylinder in a wind tunnel.
As a natural and simple measure of wind gusts we use the statistics of {\it velocity increments} $\delta u_{\tau}$ of the fluctuations:
\begin{equation}
	\delta u_{\tau}\;:=\;u(t+\tau)-u(t) \;\;\; .
	\label{increments}
\end{equation}
The increments directly measure the velocity difference after a characteristic time $\tau$ (illustrated in Fig.\ref{trapez}). So a high increment exceeding a certain threshold $S$ ($\delta u_{\tau}>S$) can be defined as a gust.\\
\noindent For a statistical analysis we are interested in how frequent a certain increment value occurs and whether this frequency depends on $\tau$. Therefore we calculate the {\it probability density functions} (pdfs) $P(\delta u_{\tau})$ of the increments. In Fig. \ref{windpdf} the pdfs for 5 different values of $\tau$ are shown. These distributions are all characterized by marked fat {\it tails} and a peak around the mean value. Such pdfs are called {\it intermittent} and differ extremely from a Gaussian distribution that is commonly considered to be the suitable distribution for continuous random processes.\\
A {\it normal distribution} is uniquely defined by its mean value $\mu$  and its standard deviation $\sigma$. Thus every distribution can be compared to a normal distribution in a quantitative way. In Fig. \ref{gauss} we compare one of the measured pdfs ($\tau=4\;s$) with a Gaussian distribution with the same $\sigma$. In this presentation the different behaviour of the tails of both distributions becomes evident.
Note that the high increments of the wind pdfs -- located in the tails -- correspond to strong  gusts. For instance the value of
$\delta u_{\tau}=7\cdot \sigma$ corresponds to a velocity ascending of $5.6\;m/s$ during $4\;s$). As shown in Fig. \ref{gauss} the measured probability of our wind data is about $10^{6}$ times higher than for a Gaussian distribution with the same standard deviation! This case is represented by the arrow in Fig. \ref{gauss}. The value $10^{6}$ -- for instance -- means that a certain gust which is observed about five times a day should be observed just once in 500 years if the distribution were a Gaussian instead of the observed intermittent one. \\
But intermittent distributions seem to appear quite often in natural or economical systems like in earthquake-(\cite{schertz}), foreign exchange market- (\cite{peinke}) or even in some traffic-statistics (\cite{vasi}).
What kind of statistics do we get in the case of local isotropic and stationary laboratory experiments?
The typical probability density functions in laboratory turbulence -- as shown in Fig. \ref{Vergleich} -- change from intermittent ones for small values of $\tau$ to rather Gaussian shaped distributions with increasing $\tau$. When $\tau\approx T$ with $T$ being the {\it correlation time} (integral time) the approach to a normal distribution is marked:
\begin{equation}
	T \;=\; \int \limits_{0}^{\infty} R(\tau)d\tau\;\;\; .
	\label{korr}
\end{equation}
$R(\tau)$ is the correlation function of the fluctuations. In our case the correlation time  of the atmospheric flow is $34\;s$ and $6\; ms$ for the laboratory data \cite{com}.\\
Note that for the pdfs of the atmospheric velocity field this characteristic change of shape, even for $\tau$-values higher than $T$ (as shown in Fig. \ref{windpdf}) is not observed.\\ 
 As already mentioned a fundamental difference between atmospheric and laboratory turbulence is that the latter is stationary. In laboratory experiments one usually deals with a fixed value and direction of the mean wind speed $\bar{u}$, which is obviously never the case for atmospheric wind fields. Therefore in a second step we calculate the pdfs of the atmospheric increments only for certain mean velocity intervals. That means that only those increments are taken into account with $\bar{u}$ ranging in a narrow velocity interval with a width of typically $1\;m/s$. These {\it conditioned} pdfs $P(\delta u_{\tau}|\bar{u})$
show a similar qualitative change of shape like those of the laboratory experiment\footnote{Only for very small values of $\bar{u}$ ($u<1\;m/s$) this change of shape is not observed.} which is illustrated in Fig. \ref{Vergleich} b). \\
To quantify this similarity we use a well established fit by an empirical explicite function for the pdf. This formula was derived in \cite{castaing} on the basis of Kolmogorov's understanding of a turbulent cascade:
\begin {eqnarray}
	P(\delta u_{\tau}) &=& \frac{1} { 2\pi\lambda_{\tau} }\int 
	\limits_{0}^{\infty} exp( -\frac{\delta u_{\tau}^{2} } {
	2s^{2} }) \cdot exp(-\frac{ln^{2}(s/s_{0})} {2\lambda_{\tau}^{2} } ) 
	\frac{d(lns)}{s}\;\;\; .
\label{c7}
\end{eqnarray}

\begin{figure}[H]
  \begin{center}
    \epsfig{file=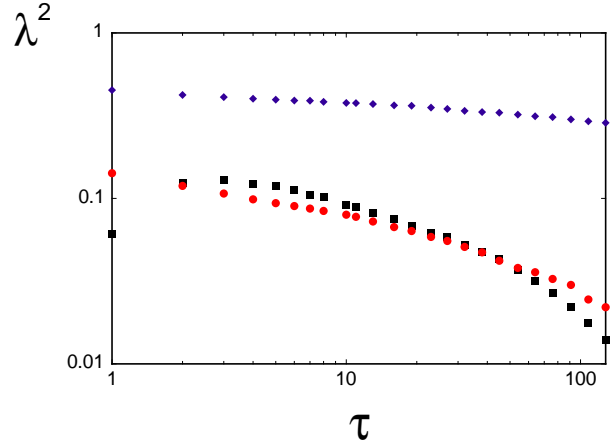, width=8.0cm}
  \end{center}
 	\caption{\it $\lambda_{\tau}^{2}$ as a function of the increment distance $\tau$ is 	shown for (from top to 	bottom) the unconditioned, the conditioned and the laboratory 	increments. For the last $\tau=1$ means $10^{-5}\;s$ for the two former $0.25\;s$   	For a  clearer visualisation a double-logarithmic presentation is used.}
 	\label{lama}
\end{figure}

\noindent In Fig. \ref{windpdf} and \ref{Vergleich}  these fitfunctions are represented by the solid lines.
$\lambda_{\tau}^{2}$ is the fundamental parameter ({\it formparameter}) in equation (\ref{c7}) and determines the shape of the probability distribution. As it can easily be seen equation (\ref{c7}) reduces to a Gaussian distribution if $\lambda_{\tau}^{2}$ is zero:
\begin{equation}
	\lim_{\lambda_{\tau}^{2} \to o} P(\delta u_{\tau})\;=\; 	\frac{1}{s_{0}\sqrt{2\pi}}exp(-\frac{\delta 	u_{\tau}^{2}}{2s_{0}^{2}})\;\;\; .
\label{gausi}
\end{equation}\\
On the other hand the more $\lambda_{\tau}^{2}$ increases the more intermittent the distributions become. In this way the parameter $\lambda_{\tau}^{2}$ may serve to compare the pdfs with each other in a more quantitative way. In Fig. \ref{lama} the evolution of the formparameter as a function of the increment distance $\tau$ is shown. \\
Other laboratory measurements (see \cite{castaing,chabaud}) of $\lambda_{\tau}^{2}$ have shown evidence that it saturates approximately at $0.2$. As shown in Fig. \ref{lama} the formparameter of the conditioned wind increments as well as of the laboratory ones is approximately $0.2$ for small $\tau-$values. Furthermore  it tends to zero with increasing $\tau$. None of these two features is observed in the case of the unconditioned increments, $\lambda_{\tau}^{2}$ is rather independent from $\tau$ with a value of about $0.7$. A constant behaviour of $\lambda_{\tau}^{2}$ means that the form of the pdfs remain unchanged, while its variance may change. Together with the power law scaling of $<\delta u_{\tau}^{2}\approx r^{2/3}>$ we get the unusual result, that for pdfs with constant intermittent shapes the scaling of the n-th moments $<\delta u_{\tau}^{n}\approx r^{\xi_{n}}>$ follow the proposed law of Kolmogorov 41\cite{kol}, $\xi_{n}=n/3$, i.e. without intermittency corrections for the exponents.\\
Thus we have shown that the anomalous statistics of wind fluctuations on discrete time intervals -- which are obviously related to wind gusts -- can be reduced to the well known intermittent (anomalous) statistics of local isotropic turbulence. This result deviates from results of wind data reported in \cite{kantz}, where it is claimed that their unconditioned wind pdfs behave like stationary ones from laboratory measurements. Thus on the basis of our stochastic analysis we claim that wind gusts are no independent structures but are a part of small scale turbulence.\\

\section{Waiting time distribution}

\begin{figure}[H]
  \begin{center}
    \epsfig{file=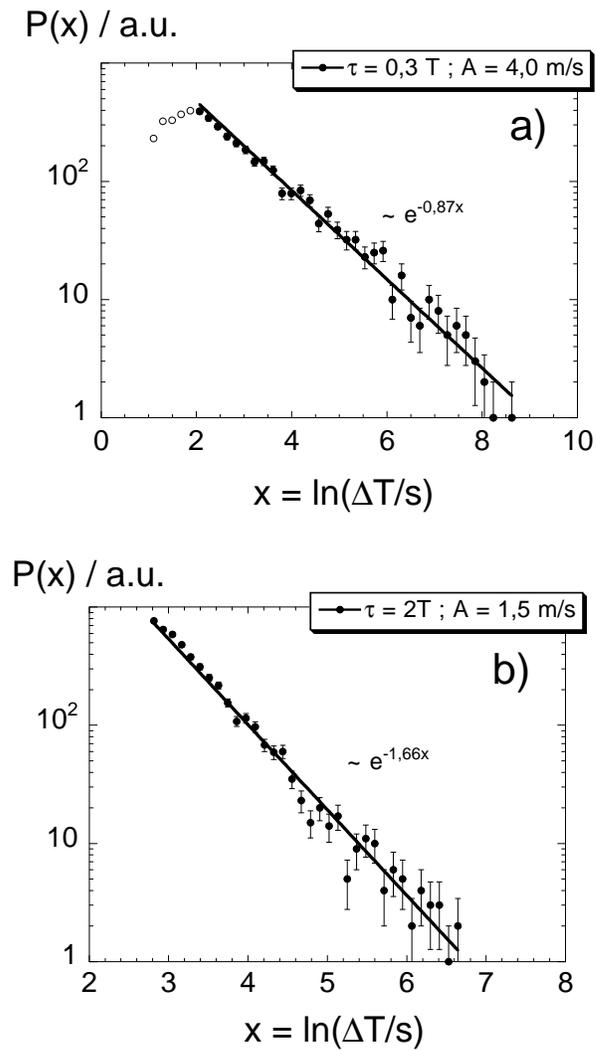, width=8.0cm}
  \end{center}
 	\caption{\it  The filled symbols illustrate the distribution of the logarithm $x$ of the 	waiting times $\Delta T$ between 	successive gusts. Additionally the fitfunctions are 	drawn in (solid lines). $x=8$ for instance corresponds to a time distance of $\Delta 	T \approx 1000\;s$.}
 	\label{wtime}
\end{figure}

So far we have shown how the atmospheric turbulence is related to the laboratory one in a statistical way. This probabalistic ansatz describes the frequency with which certain gusts occur but it is not clear how they are distributed in time. In this sense we now examine the inter-event-times between successive wind gusts.\\
The marked fat tail behaviour of the unconditioned pdfs -- as illustrated in Fig. \ref{windpdf} a) -- points at an 
interesting effect. In \cite{love} the 
equivalence between the divergence of the moments $<x^{q}>$ and the 
hyperbolic (intermittent) form of pdfs which leads to a power law 
behaviour of the probability distribution is emphasized:
\begin{equation}
	p(x\geq S)\propto S^{-q}	\;\;\; , S>>1\;\;\;.
	\label{schwelle}
\end{equation}
A famous example of such a natural power law behaviour is the {\it 
Gutenberg-Richter-law} \cite{richter} that describes the 
frequency N of earthquakes with a magnitude being greater than a 
certain threshold M (magnitude):  
\begin{eqnarray}
	log(N(E)) &=& a-bM \nonumber\\ \Leftrightarrow N(E) &\propto& 
	E^{-b}\;\;\; 	.
	\label{richter}
\end{eqnarray}
$N(E)$ describes the number of occurances of those earthquakes with an energy output higher than a threshold $E_{0}$. The magnitude M is proportional to the logarithm of $E$. The same result is found for the energy flux distribution of a wind field 
\cite{schertz}. But also the waiting time distribution of fore- and after shocks obey 
a power-law, what is known as the {\it Omori-law} \cite{omori}.\\
In this sense we now examine the waiting time distribution of wind gusts. Therefore 
we refer to the idealized gust represented in Fig. \ref{trapez} 
choosing different thresholds $A$ and different increment 
distances $\tau$ (see eq. \ref{increments}). Always when the condition $\delta u_{\tau}>A$ is fulfilled a gust event is registered. To avoid that one event is counted several times we use the condition that the distance 
between two successive events $\Delta T$ is at least $\tau$. Due to the cumulation of occurances for small time distances we choose a logarithmic time.\\
The distributions for $A=4.0\;m/s$ and
$\tau=0.3\;T$ (about ten seconds) and for $A=1.5\;m/s$ and $\tau=2.0\;T$ are shown in Fig. \ref{wtime} a) respectively \ref{wtime} b).
It can easily be seen that the exponential fit in Fig. \ref{wtime} corresponds to a power 
law concerning the waiting times $\Delta T$. Only the exponent seems to depend on $A$ and $\tau$.\\
Considering
\begin{equation}
	x=ln(\Delta T) \;\;\; 
\end{equation}
and the normalisation
\begin{equation}
	1 = \int \limits_{0}^{\infty} P(x)dx = \int \limits_{0}^{\infty} P(x(\Delta 	t))\frac{dx}{d(\Delta T)}d(\Delta T)\;\;\; ,
\end{equation}
the used fitfunction
\begin{equation}
	P(x) \propto e^{-cx} 
\end{equation}
can also be written as a function of $\Delta T$ which leads to a power-law:
\begin{eqnarray}
	P(\Delta T) & \propto & k\cdot e^{(ln(\Delta T))^{-c}}\cdot \Delta 
	T^{-1} \nonumber \\
	& \propto & k \cdot (\Delta T)^{-c-1} \;\;\; .
\end{eqnarray}
This power law behaviour of the waiting time distributions is only observed for the atmospheric wind data and not for the stationary laboratory one. \\
\clearpage

\section{Summary}

On the basis of well defined velocity increments an analogous analysis of measured wind data and measured data from a turbulent wake was performed. The statistics of velocity increments, as a related statistics to the occurance frequency of wind gusts, showed that they are highly intermittent. These anomalous (not Gaussian distributed) statistics explain an increased high probability of finding strong gusts. This could be set in analogy with turbulence measurements of an idealized, local isotropic laboratory flow if a proper condition on a mean wind speed was performed. So far the statistics of wind gusts can be interpreted as a superposition of idealized packets of turbulence at different Reynoldsnumbers.\\
As a further statistical feature of wind gusts we have investigated the waiting times between successive gusts exceeding a certain strength. Here we find power-law-statistics (fractal statistics)  -- similar to earthquake statistics -- that can not be reproduced in laboratory measurements. \\
To conclude we have shown two important aspects of wind gusts. The overall occurance statistics could be set into analogy to the anomalous statistics of velocity increments in local isotropic turbulence. 
The time structure of successive gust events displays fractal behaviour.


\begin{thebibliography} {17}
\bibitem{com}			For the wind data the upper limit of the integral was taken to be 							$30\;min$.
\end{thebibliography}
\end{document}